\documentclass[aps,amsmath,preprint,epsf,superscriptaddress,nofootinbib]{revtex4-1}
\usepackage{graphicx}
\usepackage{bm}
\usepackage{color}
\usepackage{amsmath,amssymb}	
\usepackage{hyperref}
\usepackage{setspace}

\begin{document}
\def\a{\alpha}
\def\b{\beta}
\def\c{\varepsilon}
\def\d{\delta}
\def\e{\epsilon}
\def\f{\phi}
\def\g{\gamma}
\def\h{\theta}
\def\k{\kappa}
\def\l{\lambda}
\def\m{\mu}
\def\n{\nu}
\def\p{\psi}
\def\q{\partial}
\def\r{\rho}
\def\s{\sigma}
\def\t{\tau}
\def\u{\upsilon}
\def\v{\varphi}
\def\w{\omega}
\def\x{\xi}
\def\y{\eta}
\def\z{\zeta}
\def\D{{\mit \Delta}}
\def\G{\Gamma}
\def\H{\Theta}
\def\L{\Lambda}
\def\F{\Phi}
\def\P{\Psi}

\def\S{\Sigma}

\def\o{\over}
\def\beq{\begin{eqnarray}}
\def\eeq{\end{eqnarray}}
\newcommand{\gsim}{ \mathop{}_{\textstyle \sim}^{\textstyle >} }
\newcommand{\lsim}{ \mathop{}_{\textstyle \sim}^{\textstyle <} }
\newcommand{\vev}[1]{ \left\langle {#1} \right\rangle }
\newcommand{\bra}[1]{ \langle {#1} | }
\newcommand{\ket}[1]{ | {#1} \rangle }
\newcommand{\EV}{ {\rm eV} }
\newcommand{\KEV}{ {\rm keV} }
\newcommand{\MEV}{ {\rm MeV} }
\newcommand{\GEV}{ {\rm GeV} }
\newcommand{\TEV}{ {\rm TeV} }
\def\diag{\mathop{\rm diag}\nolimits}
\def\Spin{\mathop{\rm Spin}}
\def\SO{\mathop{\rm SO}}
\def\O{\mathop{\rm O}}
\def\SU{\mathop{\rm SU}}
\def\U{\mathop{\rm U}}
\def\Sp{\mathop{\rm Sp}}
\def\SL{\mathop{\rm SL}}
\def\tr{\mathop{\rm tr}}

\def\IJMP{Int.~J.~Mod.~Phys. }
\def\MPL{Mod.~Phys.~Lett. }
\def\NP{Nucl.~Phys. }
\def\PL{Phys.~Lett. }
\def\PR{Phys.~Rev. }
\def\PRL{Phys.~Rev.~Lett. }
\def\PTP{Prog.~Theor.~Phys. }
\def\ZP{Z.~Phys. }

\newcommand{\drawsquare}[2]{\hbox{%
\rule{#2pt}{#1pt}\hskip-#2pt
\rule{#1pt}{#2pt}\hskip-#1pt
\rule[#1pt]{#1pt}{#2pt}}\rule[#1pt]{#2pt}{#2pt}\hskip-#2pt
\rule{#2pt}{#1pt}}

\def\vbr{\vphantom{\sqrt{F_e^i}}}
\newcommand{\fund}{\drawsquare{6.5}{0.4}}
\newcommand{\afund}{\overline{\fund}}
\newcommand{\symm}{\drawsquare{6.5}{0.4}\hskip-0.4pt%
        \drawsquare{6.5}{0.4}}
\newcommand{\asymm}{\raisebox{-3pt}{\drawsquare{6.5}{0.4}\hskip-6.9pt%
        \raisebox{6.5pt}{\drawsquare{6.5}{0.4}}}}
\newcommand{\asymmthree}{\raisebox{-7pt}{\drawsquare{6.5}{0.4}}\hskip-6.9pt%
\raisebox{-0.5pt}{\drawsquare{6.5}{0.4}}\hskip-6.9pt%
\raisebox{6pt}{\drawsquare{6.5}{0.4}}}
\newcommand{\asymmfour}{\raisebox{-10pt}{\drawsquare{6.5}{0.4}}\hskip-6.9pt%
\raisebox{-3.5pt}{\drawsquare{6.5}{0.4}}\hskip-6.9pt%
\raisebox{3pt}{\drawsquare{6.5}{0.4}}\hskip-6.9pt%
        \raisebox{9.5pt}{\drawsquare{6.5}{0.4}}}
\newcommand{\Ythrees}{\raisebox{-.5pt}{\drawsquare{6.5}{0.4}}\hskip-0.4pt%
          \raisebox{-.5pt}{\drawsquare{6.5}{0.4}}\hskip-0.4pt%
          \raisebox{-.5pt}{\drawsquare{6.5}{0.4}}}
\newcommand{\Yfours}{\raisebox{-.5pt}{\drawsquare{6.5}{0.4}}\hskip-0.4pt%
          \raisebox{-.5pt}{\drawsquare{6.5}{0.4}}\hskip-0.4pt%
          \raisebox{-.5pt}{\drawsquare{6.5}{0.4}}\hskip-0.4pt%
          \raisebox{-.5pt}{\drawsquare{6.5}{0.4}}}
\newcommand{\Ythreea}{\raisebox{-3.5pt}{\drawsquare{6.5}{0.4}}\hskip-6.9pt%
        \raisebox{3pt}{\drawsquare{6.5}{0.4}}\hskip-6.9pt
        \raisebox{9.5pt}{\drawsquare{6.5}{0.4}}}
\newcommand{\Yfoura}{\raisebox{-3.5pt}{\drawsquare{6.5}{0.4}}\hskip-6.9pt%
        \raisebox{3pt}{\drawsquare{6.5}{0.4}}\hskip-6.9pt
        \raisebox{9.5pt}{\drawsquare{6.5}{0.4}}\hskip-6.9pt
        \raisebox{16pt}{\drawsquare{6.5}{0.4}}}
\newcommand{\Yadjoint}{\raisebox{-3.5pt}{\drawsquare{6.5}{0.4}}\hskip-6.9pt%
        \raisebox{3pt}{\drawsquare{6.5}{0.4}}\hskip-0.4pt
        \raisebox{3pt}{\drawsquare{6.5}{0.4}}}
\newcommand{\Ysquare}{\raisebox{-3.5pt}{\drawsquare{6.5}{0.4}}\hskip-0.4pt%
        \raisebox{-3.5pt}{\drawsquare{6.5}{0.4}}\hskip-13.4pt%
        \raisebox{3pt}{\drawsquare{6.5}{0.4}}\hskip-0.4pt%
        \raisebox{3pt}{\drawsquare{6.5}{0.4}}}
\newcommand{\Yflavor}{\Yfund + \overline{\Yfund}} 
\newcommand{\Yoneoone}{\raisebox{-3.5pt}{\drawsquare{6.5}{0.4}}\hskip-6.9pt%
        \raisebox{3pt}{\drawsquare{6.5}{0.4}}\hskip-6.9pt%
        \raisebox{9.5pt}{\drawsquare{6.5}{0.4}}\hskip-0.4pt%
        \raisebox{9.5pt}{\drawsquare{6.5}{0.4}}}%


\preprint{IPMU16-0071}
\preprint{CTPU-16-14}
\bigskip

\title{Thermal Relic Dark Matter Beyond the Unitarity Limit}

\author{Keisuke Harigaya}
\email[e-mail: ]{keisukeharigaya@berkeley.edu}
\affiliation{Berkeley Center for Theoretical Physics, Department of Physics,
University of California, Berkeley, CA 94720, USA}
\affiliation{Theoretical Physics Group, Lawrence Berkeley National Laboratory, Berkeley, CA 94720, USA}

\author{Masahiro Ibe}
\email[e-mail: ]{ibe@icrr.u-tokyo.ac.jp}
\affiliation{Kavli IPMU (WPI), UTIAS, The University of Tokyo, Kashiwa, Chiba 277-8583, Japan}
\affiliation{ICRR, The University of Tokyo, Kashiwa, Chiba 277-8582, Japan}

\author{Kunio Kaneta}
\email[e-mail: ]{kaneta@ibs.re.kr}
\affiliation{Center for Theoretical Physics of the Universe,
Institute for Basic Science (IBS), Daejeon 34051, Republic of Korea}

\author{Wakutaka Nakano}
\email[e-mail: ]{m156077@icrr.u-tokyo.ac.jp}
\affiliation{Kavli IPMU (WPI), UTIAS, The University of Tokyo, Kashiwa, Chiba 277-8583, Japan}
\affiliation{ICRR, The University of Tokyo, Kashiwa, Chiba 277-8582, Japan}

\author{Motoo Suzuki}
\email[e-mail: ]{m0t@icrr.u-tokyo.ac.jp}
\affiliation{Kavli IPMU (WPI), UTIAS, The University of Tokyo, Kashiwa, Chiba 277-8583, Japan}
\affiliation{ICRR, The University of Tokyo, Kashiwa, Chiba 277-8582, Japan}

\date{\today}

\begin{abstract}
We discuss a simple  model of thermal relic dark matter 
whose mass can be much larger than the so-called unitarity limit on the mass of point-like particle dark matter.
The model consists of new strong dynamics with one flavor of 
fermions in the fundamental representation which is much heavier than the dynamical scale 
of the new strong dynamics.
Dark matter is identified with the lightest baryonic hadron of the new dynamics.
The baryonic hadrons annihilate into the mesonic hadrons of the new strong dynamics
when they have large radii.
Resultantly,  thermal relic dark matter with a mass in the PeV range is possible. 
\end{abstract}

\maketitle

\section{Introduction}
Despite overwhelming evidence of the existence of dark matter, its identity has remained unknown 
for almost eighty years since its first postulation. 
We are only almost certain that dark matter is not a part of the standard model of the elementary particle physics. 
Therefore,  it is one of the most important tasks of modern particle physics to identify
the origin of dark matter (see e.g. \cite{Bertone:2004pz,Murayama:2007ek,Feng:2010gw}).

Among various candidates for dark matter, thermal relic dark matter is one of the most attractive 
candidates~\cite{Lee:1977ua,Hut:1977zn,Wolfram:1978gp,Srednicki:1988ce,Kolb:1990vq,Gondolo:1990dk}. 
The thermal relic dark matter explains the observed dark matter density by its freeze-out from the thermal bath. 
For the $s$-wave annihilation, for example, the observed dark matter density is reproduced when
the annihilation cross section satisfies $\langle \sigma v\rangle \simeq 3\times10^{-26}$\,cm$^3/$s.
The beauty of  thermal relic dark matter is that the resultant density does not depend on 
the initial condition as long as dark matter was in the thermal equilibrium in the early universe. 

As an important consequence of thermal relic dark matter, 
there is an upper limit on the mass of dark matter from the so-called unitarity limit
on the annihilation cross section~\cite{Griest:1989wd}.
In fact, the $s$-wave annihilation cross section of dark matter with a mass $M$ is limited from above by unitarity;
\begin{eqnarray}
\sigma v \lesssim \frac{4\pi}{M^2 v}\ .
\end{eqnarray}
Combined with the required cross section mentioned above, the upper limit on the dark matter mass turns out to be
about a hundred TeV.

In this paper, we challenge the unitarity limit on the mass of thermal relic dark matter.
In fact, the above unitarity limit applies when the dark matter is a point-like particle.
If dark matter is a bound state with a large radius compared with its Compton length,  
on the other hand, it may have a geometrical cross section for 
annihilation~\cite{Griest:1989wd} (see also \cite{Nardi:1990ku,Kang:2006yd,Huo:2015nwa}).
With the larger cross section,  thermal relic dark matter with a mass much larger than a few hundred TeV is possible.
We construct a simple model where bound state dark matter annihilate while they have large radii and hence have a large geometrical cross section.

This mechanism should be compared with the enhancement of the dark matter annihilation cross section
by the so-called Sommerfeld enhancement~\cite{Hisano:2003ec,Hisano:2004pv,Hisano:2004ds,Hisano:2006nn}.
In this case, the dark matter itself is a rather point-like particle, and hence, the enhanced 
cross section does satisfy the unitarity limit of point-like particles (see Ref.~\cite{vonHarling:2014kha} for recent discussion).%
\footnote{The same is true in the models with the so-called Breit-Wigner enhancement~\cite{Ibe:2008ye,Ibe:2009mk}.} 
In a model discussed in this paper, on the other hand, dark matter itself is a bound state 
and has an annihilation cross section of a geometrical size with which the number 
density of  dark matter is significantly reduced.

The organization of this paper is as follows.
In section\,\,\ref{sec:model}, we introduce a model based on a simple strongly coupled gauge theory.
In section\,\,\ref{sec:relic}, we discuss thermal history and the relic density of dark matter.
The final section is devoted to conclusions and discussions. 
There, we also comment on a possible application of the present model to explain the excess 
of the observed flux of extraterrestrial neutrinos
in the PeV range~\cite{Aartsen:2013jdh,Aartsen:2015zva,Aartsen:2015knd}.
In the appendices, we also discuss two alternative models.

\section{Model of Dark Matter with Axion Portal}
\label{sec:model}
Let us consider an $SU(N_c)$ gauge theory with one-flavor of Weyl fermions, ($U$, $\bar{U}$),
in the fundamental and the anti-fundamental representations.
We call ($U,\bar{U}$) the quarks in the following.
The quark does not carry any gauge charges under the Standard Model gauge groups.
For a while, we assume that the quark possesses a mass, $M_U$.

As a special feature of the present model, we arrange the dynamical scale of $SU(N_c)$, $\Lambda_{\rm dyn}$, 
to be much smaller than $M_U$. 
That is, we take the gauge coupling constant at the renormalization scale around $M_U$ small;
\begin{eqnarray}
\label{eq:aNc}
\alpha_{N_c}(M_U)  =  \left(\frac{1}{2\pi} \left(\frac{11}{3}N_c\right)\log\frac{M_U}{\Lambda_{\rm dyn}}\right)^{-1} \simeq {\cal O}(0.1) \times \frac{1}{N_c} \ ,
\end{eqnarray}
where $\alpha_{N_c} = g_{N_c}^2/4\pi $ is the fine-structure constant.
Below $M_U$, the model behaves as the pure-Yang Mills theory.
According to the standard understanding of QCD, this theory also exhibits confinement, 
which has been confirmed by lattice simulations e.g.~\cite{Morningstar:1997ff,Morningstar:1999rf,Lucini:2010nv} 
(see also Ref.~\cite{Gupta:1981ve} for earlier disucssion).
After confinement, the gluons of $SU(N_c)$ gauge theory are bounded into light glueballs, ${\cal S}$'s, with masses of ${\cal O}(\Lambda_{\rm dyn})$. 
The heavy quarks are, on the other hand, trapped into quarkonia (we call mesons, ${\cal M}$'s) or in 
 heavy baryons, ${\cal B}$'s. 
The masses of those heavy mesons and baryons are $M_{\cal M}\simeq 2\times M_U$  and $M_{\cal B}\simeq N_c\times M_U$, respectively.

A striking feature of this setup is that the chromo-electric flux tube of $SU(N_c)$~\cite{Nambu:1974zg,tHooft:1975pu,Mandelstam:1974pi}
can be stretched much longer than $\Lambda_{\rm dyn}^{-1}$ due to the heaviness of the quarks~\cite{Kang:2008ea}.
It eventually breaks-up and creates a pair of a quark and an anti-quark when its length becomes of ${\cal O}(M_U/F_{N_c})$ 
where $F_{N_c}$ denotes the string tension made by the flux tube.
Therefore, the $SU(N_c)$ gauge dynamics leads to a rather long-range force even after confinement.

The quarks are stable and can be a dark matter candidate due to a vector-like global $U(1)$ symmetry under which the quarks are charged.
We call this symmetry the $U(1)_B$ symmetry.
The quarks, however, do not become dark matter as they are. 
As noted above, they are confined into hadrons when the temperature of the universe becomes 
lower than the critical temperature  $T_c = {\cal O}(\Lambda_{\rm dyn})$.
Below the critical temperature, the $U(1)_B$ charges of the quarks are inherited to the baryons, and the lightest baryon,
\begin{eqnarray}
{\cal B}_0 \propto \epsilon^{i_1i_2\cdots i_{N_c}} U_{i_1} U_{i_2} \cdots U_{i_{N_c}}\ ,
\end{eqnarray}
becomes dark matter eventually.%
\footnote{The lightest baryon, ${\cal B}_0$, possesses a spin $N_c/2$ due to the fermi-statistics.}
The mesons, on the other hand, do not carry the $U(1)_B$ charges and are not stable.
In fact, the ground state meson, for example, immediately decays into a pair of the glueballs as we will see shortly.

For a successful model of  thermal relic dark matter, the above dark matter sector needs to be connected 
to the Standard Model.
As an example of such connection, we here consider a model with ``axion portal".%
\footnote{In the appendices \ref{sec:Higgs portal} and \ref{sec:hyper}, we discuss models with ``higgs portal" 
and "hypercharged particle" to the 
Standard Model sector as  alternative examples.
We may also consider models with a ``vector portal" in which a dark photon connects the two sectors.}
For that purpose, we first replace the mass term of the quark with an interaction term to a singlet complex scalar field
$\phi$
\begin{eqnarray}
{\cal L}  = g\, \phi \, \bar{U} U  + h.c. \ ,
\end{eqnarray}
and assume that the model  possesses an approximate chiral symmetry, $U(1)_A$.
Here, $g$ denotes a coupling constant of ${\cal O}(1)$.
The quark obtains a mass $M_U = g \langle \phi\rangle$ when the $U(1)_A$ chiral symmetry is spontaneously  broken 
by a vacuum expectation value (VEV) of $\phi$.

At around the VEV of $\phi$, $\langle{\phi}\rangle =f_a/\sqrt{2}$, $\phi$ is decomposed into a scalar boson $\rho$ and a pseudo Nambu-Goldstone boson $a$,
\begin{eqnarray}
\phi = \frac{1}{\sqrt{2}}(f_a + \rho) e^{i a/f_a}\ .
\label{eq:axion}
\end{eqnarray}
The mass of the scalar boson $\rho$ is expected to be of ${\cal O}(f_a)$.
As we will see shortly, however, the mass of $\rho$ should be somewhat suppressed for
a successful model.
The ``axion" component $a$, on the other hand, obtains a mass from explicit breaking of the 
$U(1)_A$ symmetry.
When the explicit breaking effects are dominated by the $U(1)_A$ anomaly of $SU(N_c)$, 
the axion mass is estimated to be
\begin{eqnarray}
 m_a \sim \frac{\Lambda_{\rm dyn}^2}{f_a}\ ,
\end{eqnarray}
which is much smaller than the dynamical scale.

As a portal to the Standard Model, we introduce another vector-like quarks ($d'$, $\bar{d}'$) which are {not charged} under $SU(N_c)$ 
but are charged under the Standard Model gauge groups.%
\footnote{Here, for simplicity, we take the gauge charges of $\bar{d}'$ to be the same with those of the down-type quarks of the Standard Model. 
With this choice, $\bar{d}'$ can decay immediately via small mixings to the down-type quarks.}
Similarly to ($U,\bar{U}$), the newly introduced ($d'$, $\bar{d}'$) also couples to $\phi$ via,
\begin{eqnarray}
{\cal L}  = g'\, \phi \, \bar{d}' d'  + h.c. \ .
\end{eqnarray}
After integrating out ($d'$, $\bar{d}' $), we obtain  effective interactions of the axion to the Standard Model gauge bosons,
\begin{eqnarray}
\label{eq:anomaly}
{\cal L} = \frac{\alpha_{\rm QCD}}{8\pi}\frac{a}{f_a} G\tilde{G} + 
\frac{\alpha_{\rm QED}}{12\pi}\frac{a}{f_a} F\tilde{F}  \ , 
\end{eqnarray}
where $\alpha_{\rm QCD}$ and $\alpha_{\rm QED}$ are the fine-structure constants of QCD and QED, respectively.
The Lorentz indices of the field strengths $G$ (QCD) and $F$ (QED) should be understood.

Now, we have all the necessary components of the model of dark matter.
The relevant features for the following arguments are;
\begin{itemize}
\item $SU(N_c)$ gauge theory with one-flavor of quarks, ($U,\bar{U}$), whose mass is much larger than the dynamical scale $(M_U \gg \Lambda_{\rm dyn})$.
\item The mass of ($U,\bar{U}$) is generated as a result of spontaneous breaking of an approximate $U(1)_A$ chiral symmetry, i.e. $M_U = g\vev{\phi}$.
\item The axion associated with spontaneous breaking of an approximate chiral symmetry couples to both the dark matter sector and the Standard Model sector.
\item The $U(1)_B$ charge of the quarks are inherited to the baryons after the confinement.
\item The mesons decay immediately into glueballs and axions.
\item The glueballs decay into the axions which eventually decay into the Standard Model gauge bosons.
\end{itemize}
The scalar boson $\rho$ immediately decays into a pair of axions, 
and hence, it does not play a crucial role in the following discussion.

\begin{figure}[tbp]
	\centering
  \includegraphics[width=.8\linewidth]{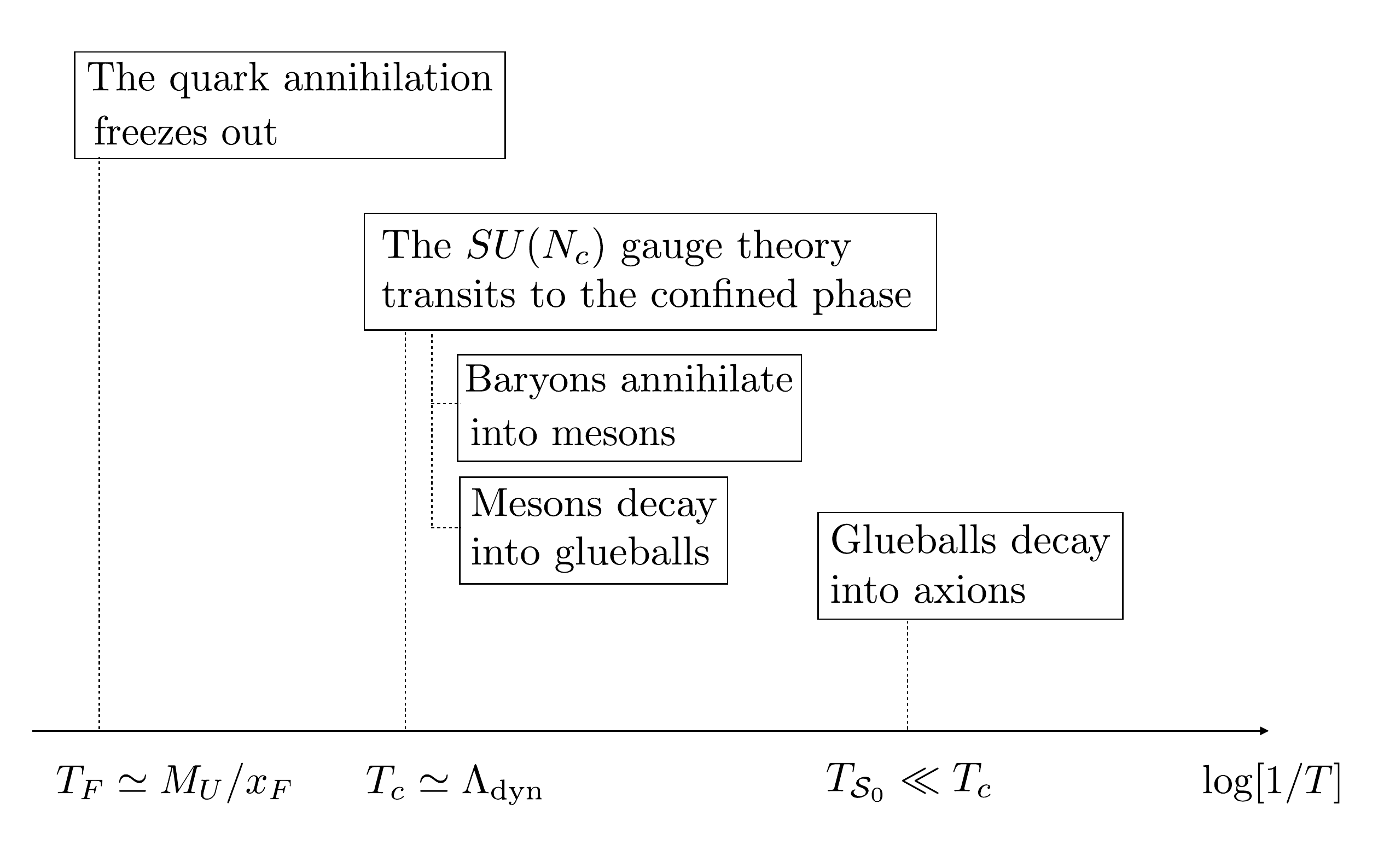}
	\caption{Summary of the thermal history of the dark matter sector. Details are discussed in the next section.
		}
	\label{fig:summary}
\end{figure}

Before closing this section, let us give a rough sketch of  thermal history  which will  be discussed in the next section.
(1) At the very early universe, the quarks $U$'s are in the thermal bath.
(2) When the temperature of the universe becomes lower than $M_U$, perturbative annihilations
of the quarks freeze-out and the relic number density of the quarks in a comoving volume is fixed.
(3) When the temperature decreases down to the critical temperature, $T_c = {\cal O}(\Lambda_{\rm dyn})$, 
the $SU(N_c)$ gauge theory exhibits confinement and the quarks are confined into either the mesons or baryons.
(4) Just below the critical temperature, most of the bound states keep large radii for a while. 
At around that time, the baryons annihilate into the mesons with a geometrical cross section,
and the number density of the baryons is significantly reduced.
Mesons, on the other hand, decay into the glueballs and axions. 
(6) Eventually, the glueballs decays into axions which in turn decay into the Standard Model 
gauge bosons.

\section{Relic Abundance of Baryonic Dark Matter}
\label{sec:relic}
\subsection{Perturbative Freeze-Out}
When the temperature of the universe is much higher than $M_U$, the quarks are in the thermal bath.
Once the temperature becomes lower than $M_U$, the annihilation process freezes-out and 
the resultant relic density per the entropy density $s$ is given by~\cite{Gondolo:1990dk},
\begin{eqnarray}
\frac{n_{U}}{s} \simeq \sqrt{\frac{45}{8\pi^2 g_*(T_F)}}\frac{x_F}{M_{\rm PL} M_U \langle\sigma_U v \rangle}\ .
\end{eqnarray}
Here, $T_F$ denotes the freeze-out temperature, $x$ the temperature mass ratio, $x = M_U/T$, $g_*(T)$
the massless degrees of freedom at $T$, and $M_{\rm PL} \simeq 2.4\times 10^{18}$\,GeV the reduced Planck scale.
The freeze-out temperature is recursively determined by,
\begin{eqnarray}
\label{eq:xF}
\ln \left[ \frac{\langle{\sigma_U v}\rangle}{2\pi^3}
\sqrt{\frac{45\pi}{g_*(T_F)}} M_{\rm PL}M_{U} g_{U} x_F^{-1/2}\right]
 = x_F\ ,
\end{eqnarray}
where $g_U$ denotes the degree of the freedom of $U$, i.e. $g_U = 4 N_c$.
A typical freeze-out temperature is given by $x_F \sim {\cal O}(10)$.

At around the freeze-out temperature, the quarks mainly annihilate into $\phi$'s, 
with the spin and color averaged annihilation cross section, 
\begin{eqnarray}
\label{eq:annihilation}
\langle{\sigma_U v}\rangle \sim \frac{1}{4N_c} \frac{\pi\alpha_g^2}{4M_U^2}\ ,
\end{eqnarray}
where $\alpha_g^2 = g^2/(4\pi)$.
We neglect the annihilation into a pair of the gluons due to Eq.\,(\ref{eq:aNc}).

Below the freeze-out temperature, the number density of the quarks are diluted by cosmic expansion,
and a typical distance between the quarks at a temperature $T$ is  given by,
\begin{eqnarray}
\label{eq:DT}
D(T) &\sim& (g_Un_{U})^{-1/3} \nonumber \\
&\sim& \frac{10^{2}}{T}\times 
\left(\frac{3}{N_c}\right)^{2/3} \left(\frac{10^6\,{\rm GeV}}{M_U}\right)^{1/3}
\left(\frac{\alpha_g}{10^{-1}}\right)^{2/3}
\left(\frac{20}{x_F}\right)^{1/3}
\left(\frac{100}{g_*(T_F)}\right)^{1/6} \ .
\end{eqnarray}
When the temperature decreases to the critical $T_c \simeq \Lambda_{\rm dyn}$,
the $SU(N_c)$ gauge interaction becomes strong and exhibits confinement. 
Below this temperature, the quarks do not freely fall separately anymore.
In the following, we discuss the fates of the bound states assuming that phase transition is first order according to Refs.\,\cite{Lucini:2003zr,Aharony:2005bq}.%
\footnote{The following arguments are not altered significantly as long as the growth of the string tension of the 
strong dynamics is fast enough.}

\subsection{Bound State Formation}
\begin{figure}[tbp]
	\centering
  \includegraphics[width=.45\linewidth]{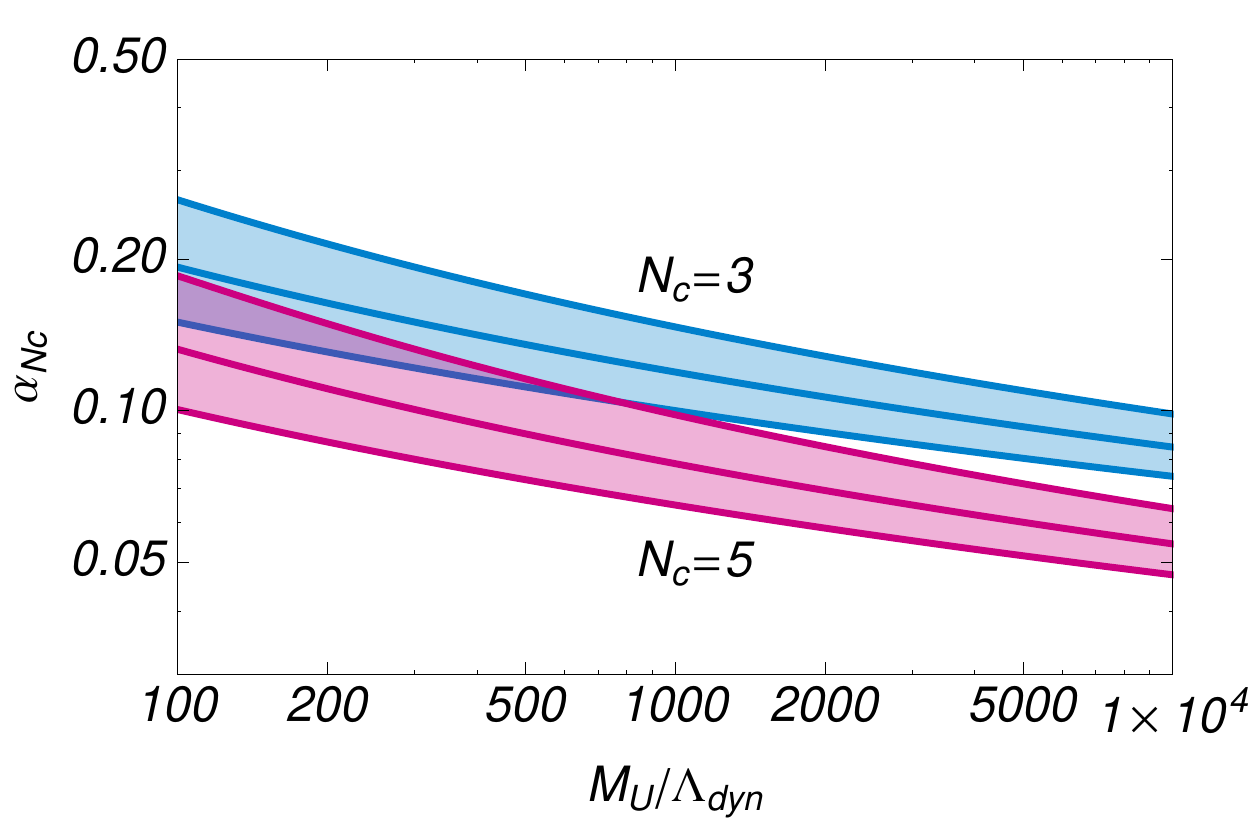}
	\caption{The coupling constant estimated at $\mu = c_\mu \k\alpha_{N_c}(\mu) M_U$ as a function of $M_U/\Lambda_{\rm dyn}$.
	In each band, we vary $c_\mu$ from $1/3$ (lower lines) to $3$ (upper lines) to show the scale dependences of the coupling constants.
	}
	\label{fig:alpNc}
\end{figure}

In order to trace the thermal history below the dynamical scale precisely, we need to 
solve the strong gauge dynamics, which is impossible with the current techniques.
Here, instead, we follow the picture in Ref.~\cite{Kang:2006yd}, and 
treat hadrons as composites of heavy quarks which 
are attracted with each other by a phenomenological potential (see e.g. \cite{Bali:2000gf}),
\begin{eqnarray}
\label{eq:potential}
V(r)  \sim -\frac{\kappa\,\alpha_{N_c}}{r} + F_{N_c}(T)\,r \ .
\end{eqnarray}
Here, $\kappa$ is an ${\cal O}(1)$ numerical factor 
that depends on the color exchanged between the quarks.
For a color singlet configuration of a quark and an anti-quark, for example, $\k = C_F = (N_c^2 - 1)/(2N_c)$.
The linear term represents the effects of non-perturbative dynamics and $F_{N_c}$ corresponds
to the tension of the flux tube.
At a high temperature, $F_{N_c}(T)$ is vanishing while $F_{N_c} \sim \Lambda_{\rm dyn}^2$
below the  critical temperature $T_c = {\cal O}(\Lambda_{\rm dyn})$.%
\footnote{The lattice simulations suggest $T_c/ \sqrt{F_{N_c}} \simeq 0.6 $
for the pure Yang-Mills $SU(N_c)$ ($N_c \ge 3$) theories\,\cite{Lucini:2003zr}.}
The gauge coupling constant $\alpha_{N_c}$ in Eq.\,(\ref{eq:En}) is, on the other hand, estimated at the renormalization scale 
corresponding to the Bohr radius $\mu \simeq \k\alpha_{N_c}(\mu) M_U$ as the leading order approximation (see Fig.\,\ref{fig:alpNc}).

When the temperature of the universe becomes lower than $T_c$, the $SU(N_c)$ gauge dynamics transits 
into the confined phase and the quarks and gulons are confined into color singlets.
In particular, the quarks at the distance $D(T_c)$ in Eq.\,(\ref{eq:DT})
are pulled with each other by the linear potential, 
and the sizes of the quark bound states become much shorter than the original distance.%
\footnote{In the parameter space we are interested in,  $D(T_c)$ is shorter than the length of the string breaking, $M_U/F_{N_c}$.
If $D(T_c) \gg M_U/F_{N_c}$, on the other hand, the strings between the quarks break up immediately and the quarks are dominantly
confined not into baryons but into mesons  especially for large $N_c$.
In this situation, the relic abundance of the baryon dark matter can be much smaller than the present scenario,
which will be discussed elsewhere.} 
It should be noted that the quarks are not accelerated even when they are pulled by the strong force
due to frictions caused by the interactions with the glueballs in the thermal bath.

\begin{figure}[tbp]
	\centering
  \includegraphics[width=.45\linewidth]{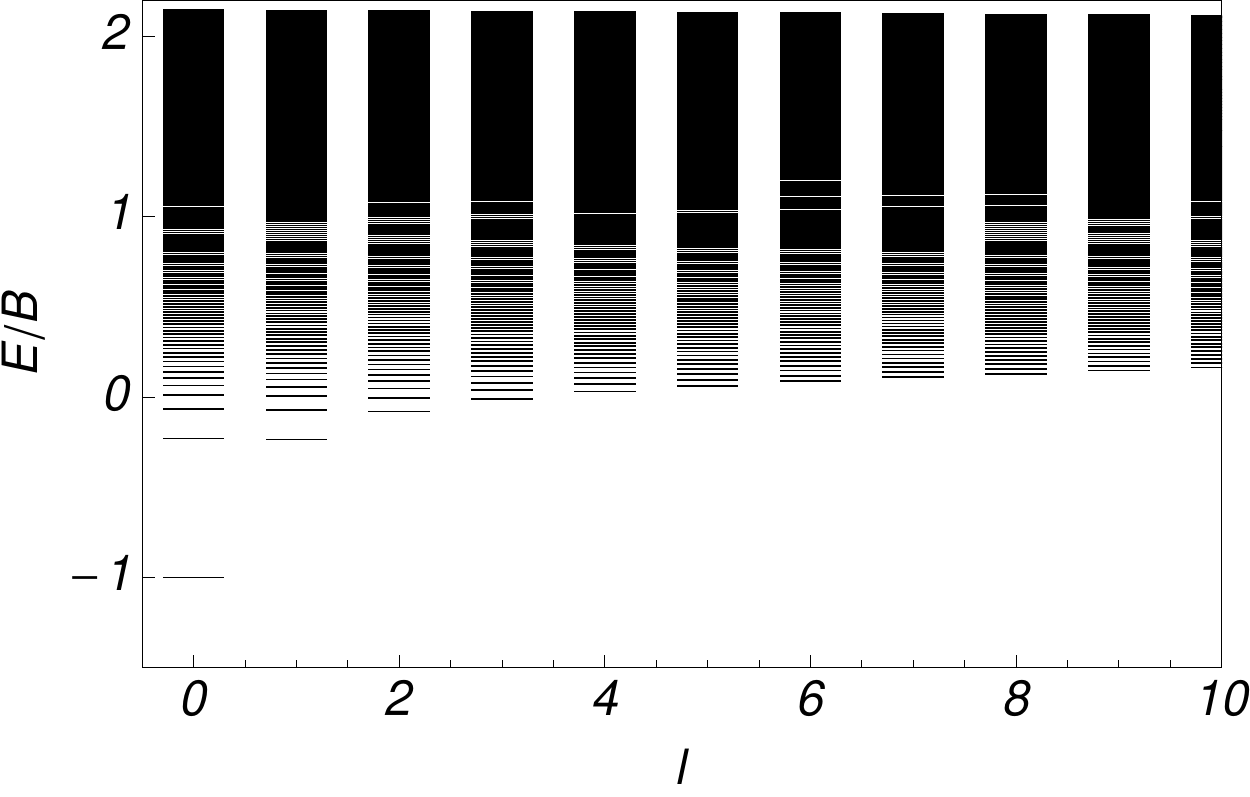}
	\caption{The approximate energy spectrum in the unit of the binding energy $B = -E_1$ 
	for the potential in Eq.\,(\ref{eq:potential}) 	as a function of the orbital angular momentum \,\cite{Hall:1984wk}. 
	Here, we take $M_U = 10^6$\,GeV, 	$\Lambda_{\rm dyn} = 10^3$\,GeV, $F_{N_c} = \Lambda_{\rm dyn}^2$,  $N_c =3$, and $\alpha_{N_c} = 0.1$. 
	}
	\label{fig:spectrum}
\end{figure}

To estimate the typical size of the bound state at a temperature, $T$,
let us consider a partition function of a quark and anti-quark bound state 
by the potential in Eq.\,(\ref{eq:potential});
\begin{eqnarray}
\label{eq:ZT}
Z[T] \simeq \sum_{n=1}^{n_{\rm max}}n^2 e^{-\left(E_n -E_1\right)/T
} + \frac{1}{(2\pi)^3}\int d^3r \, d^3p \,
 e^{-\left({p^2}/{M_U}+ F_{N_c} r - E_1 \right)/T}\ .
\end{eqnarray}
Here, the reduced mass of the two body system is given by $M_U/2$.
For the negative energy states where the Coulomb potential  is dominant, 
i.e. $r < (\kappa \alpha_{N_c}/F_{N_c})^{1/2}$,
we approximate their energy eigenvalues by 
\begin{eqnarray}
\label{eq:En}
E_{n} \simeq - \frac{\kappa^2 \alpha_{N_c}^2}{4}\frac{M_U}{n^2}\ ,\,(n\ge 1)\ .
\end{eqnarray}
Here, $n$ denotes the principal quantum number and the radii of the corresponding states are  given by,
\begin{eqnarray}
\label{eq:Bohr}
r_n \simeq \frac{2\, n^2}{\kappa\,\alpha_{N_c} M_U}\ .
\end{eqnarray}
For the positive energy states which correspond to $r > (\kappa \alpha_{N_c}/F_{N_c})^{1/2}$, on the other hand,
we approximate them by continuous spectrum (see Fig.\,\ref{fig:spectrum}).
We checked that the above approximation well reproduces a quantum statistical partition function with approximate energy eigenvalues in Ref.\,\cite{Hall:1984wk}.
For ease of the computation, we rely on the  approximation in Eq.\,(\ref{eq:ZT}) in the following arguments.

\begin{figure}[tbp]
	\centering
		\begin{minipage}{.46\linewidth}
  \includegraphics[width=\linewidth]{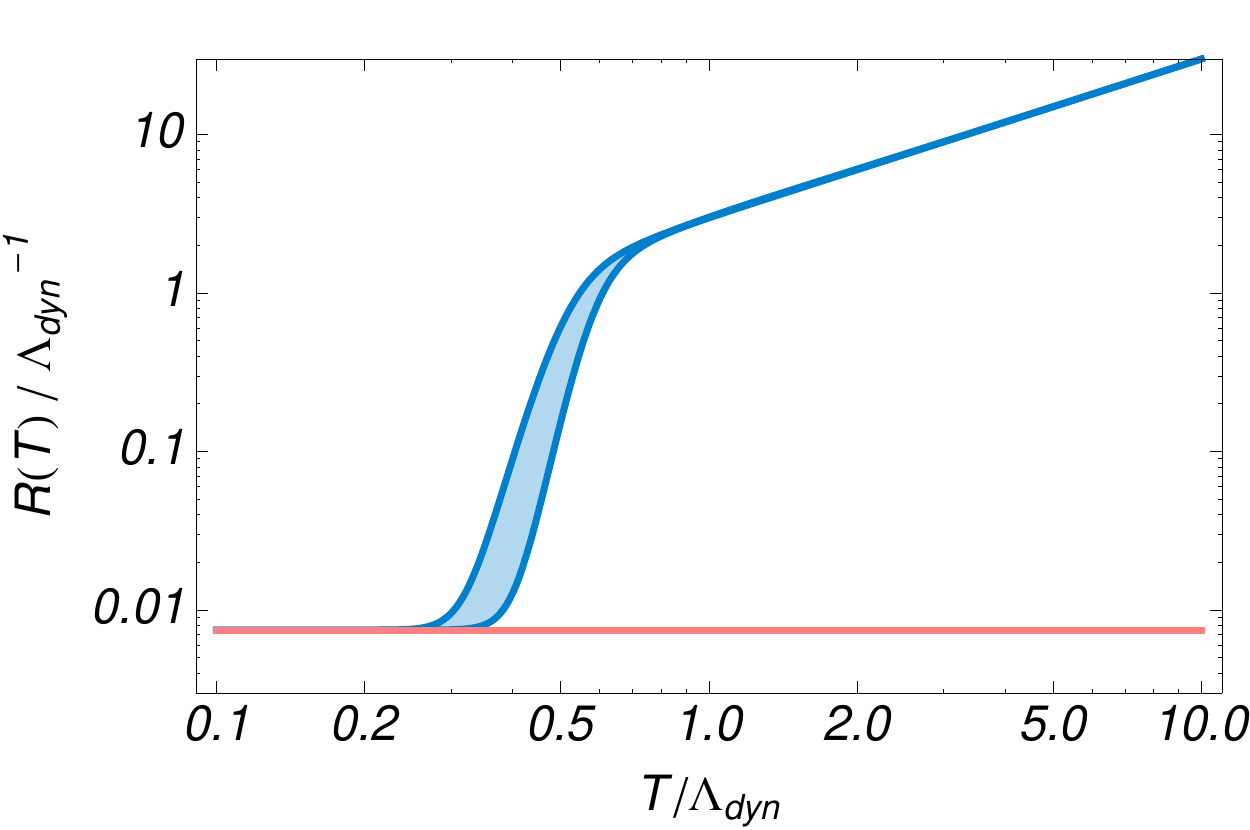}
 \end{minipage}
 \hspace{1cm}
 \begin{minipage}{.46\linewidth}
  \includegraphics[width=\linewidth]{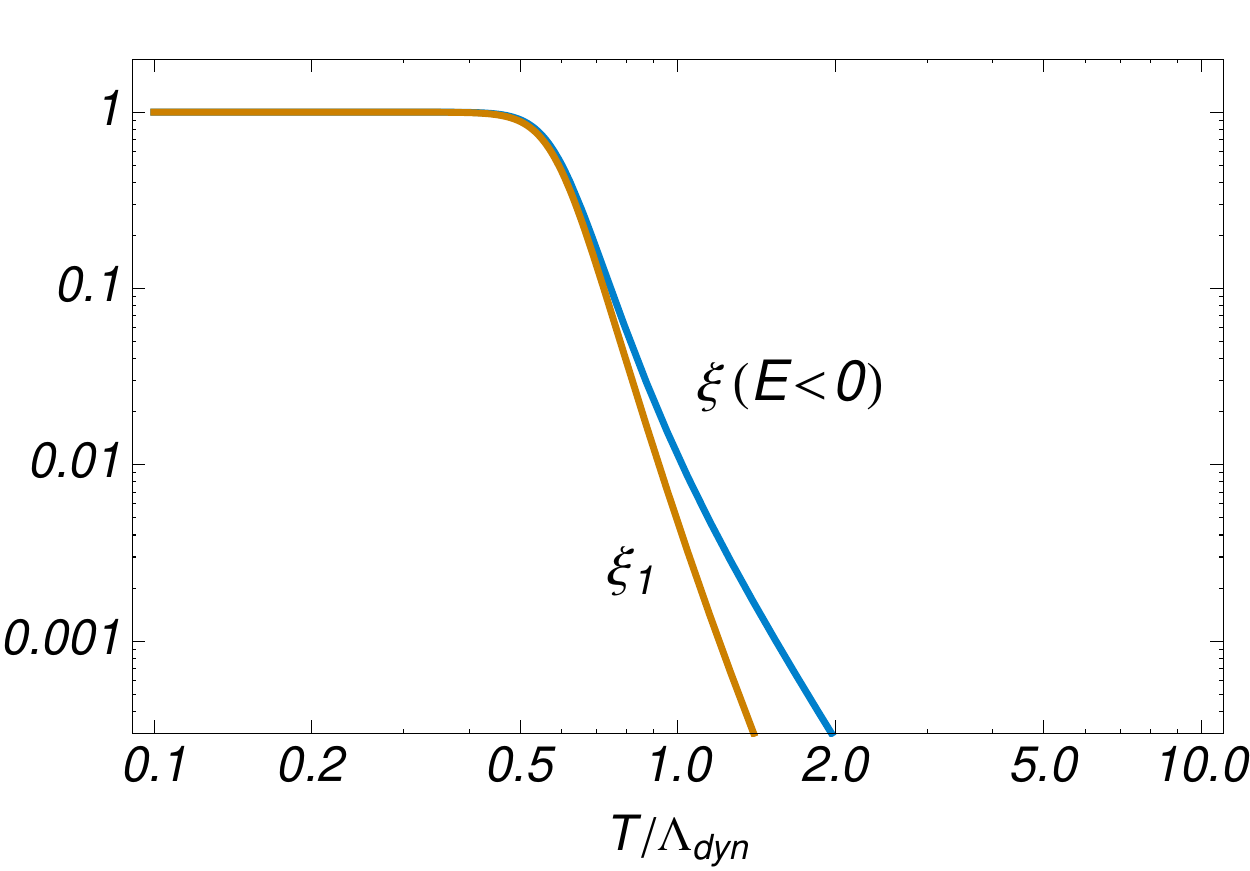}
 \end{minipage}
	\caption{(Left) A typical size of the bound states estimated by Eq.\,(\ref{eq:RT}) for $M_U = 10^6$\,GeV, $\Lambda_{\rm dyn} = 10^3$\,GeV, 
	$F_{N_c} = \Lambda_{\rm dyn}^2$, $N_c = 3$, and $\alpha_{N_c} = 0.1$.
	In the blue shaded band we vary $n_{\rm max}$ from one to three times of the one defined by $r_{n_{\rm max}} = (\kappa \alpha_{N_c}/F_{N_c})^{1/2}$.
	The horizontal red line corresponds to the Bohr radius.
	(Right) The fractional occupation numbers of the negative energy state, $\xi(E<0)$, and the ground state, $\xi_1$.
	Here, we fix $n_{\rm max}$ to be the one defined by $r_{n_{\rm max}} = (\kappa \alpha_{N_c}/F_{N_c})^{1/2}$.
	In both panels, we fix $F_{N_c}\simeq \Lambda_{\rm dyn}^2$ even for $T> T_c \simeq \Lambda_{\rm dyn}$ for presentation purpose.
	}
	\label{fig:RT}
\end{figure}

In Fig.\,\ref{fig:RT}, we show the typical size of the quark bound state for a given temperature estimated by
\begin{eqnarray}
\label{eq:RT}
R(T) &\simeq& 
\left(\sum_{n=1}^{n_{\rm max}} 
 \frac{2\, n^2}{\kappa\,\alpha_{N_c} M_U}\
 n^2 e^{-\frac{1}{T}\left(E_n -E_1\right)} \right.\nonumber \\
&&\quad\quad\quad\quad
\left. + 
 \frac{1}{(2\pi)^3}\int d^3r \, d^3p \, r\,
 e^{-\left({p^2}/{M_U}+ F_{N_c} r - E_1 \right)/T}
\right)
/{Z[T]}\ .
\end{eqnarray}
We also show the fractional occupation numbers of the negative energy state, $\xi(E<0)$, and the ground state, $\xi_1$,
\begin{eqnarray}
\label{eq:Xi}
\xi(E<0) \simeq \sum_{n=1}^{n_{\rm max}}n^2 e^{-\left(E_n -E_1\right)/T
} /Z[T]\ , \quad \quad
\xi_1 \simeq& 1 /Z[T] \ ,
\end{eqnarray}
respectively.
Here, $n_{\rm max}$ is defined by $r_{n_{\rm max}} = (\kappa \alpha_{N_c}/F_{N_c})^{1/2}$,
although the results do not depend on the precise value of $n_{\rm max}$ significantly. 
The figure shows that $R(T_c) = {\cal O}(\Lambda_{\rm dyn}^{-1})$.
Thus, we find that the bound states are in excited states below the critical temperature. 
When the temperature decreases further, the bound states are de-excited and 
the typical size becomes $r_1$ in Eq.\,(\ref{eq:Bohr}).

\begin{figure}[tbp]
	\centering
  \includegraphics[width=.45\linewidth]{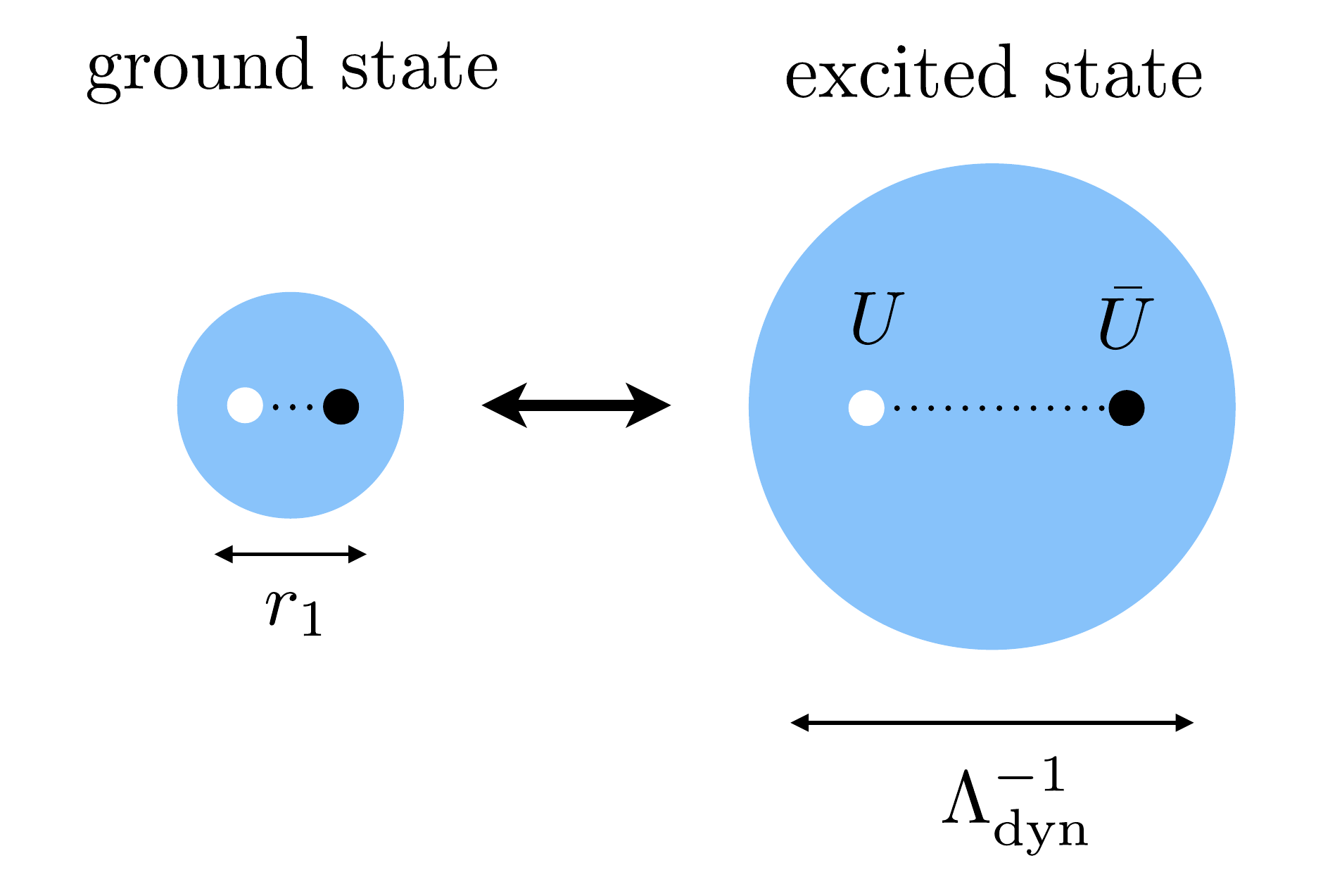}
	\caption{A schematic picture of the transition between the ground state and the excited states in the thermal bath.
	}
	\label{fig:MM}
\end{figure}

It should be noted that quarks in the ground state are knocked out to the excited states 
by  scatterings with the glueballs in the thermal bath. 
The rate of such processes is roughly given by,%
\footnote{Here, $\alpha_{N_c}$ should be estimated at around the dynamical scale and hence of ${\cal O}(1)$,
although the precise value is not relevant for our discussion.}
\begin{eqnarray}
\Gamma_{\rm ex} \sim \alpha_{N_c}^2 \left(\frac{T}{B + m_{\cal S}}\right)^2 T\, e^{-\frac{B+m_{\cal S}}{T}}\ .
\end{eqnarray}
Here $m_{\cal S}$ denotes the glueball mass which is slightly larger than the scale of the string tension
in pure Yang-Mills theories~\cite{Morningstar:1997ff,Morningstar:1999rf,Lucini:2010nv}.
In the parameter region we are interested in, $\Gamma_{\rm ex}$ is larger than the Hubble expansion rate at $T\simeq T_c$.
Therefore, the each bound state transits between the ground state to the excited states rather frequently (see Fig.\,\ref{fig:MM}).
This behavior plays a crucial role for the final dark matter abundance.

\subsection{Fate of Mesons}
\begin{figure}[tbp]
	\centering
		\begin{minipage}{.48\linewidth}
  \includegraphics[width=\linewidth]{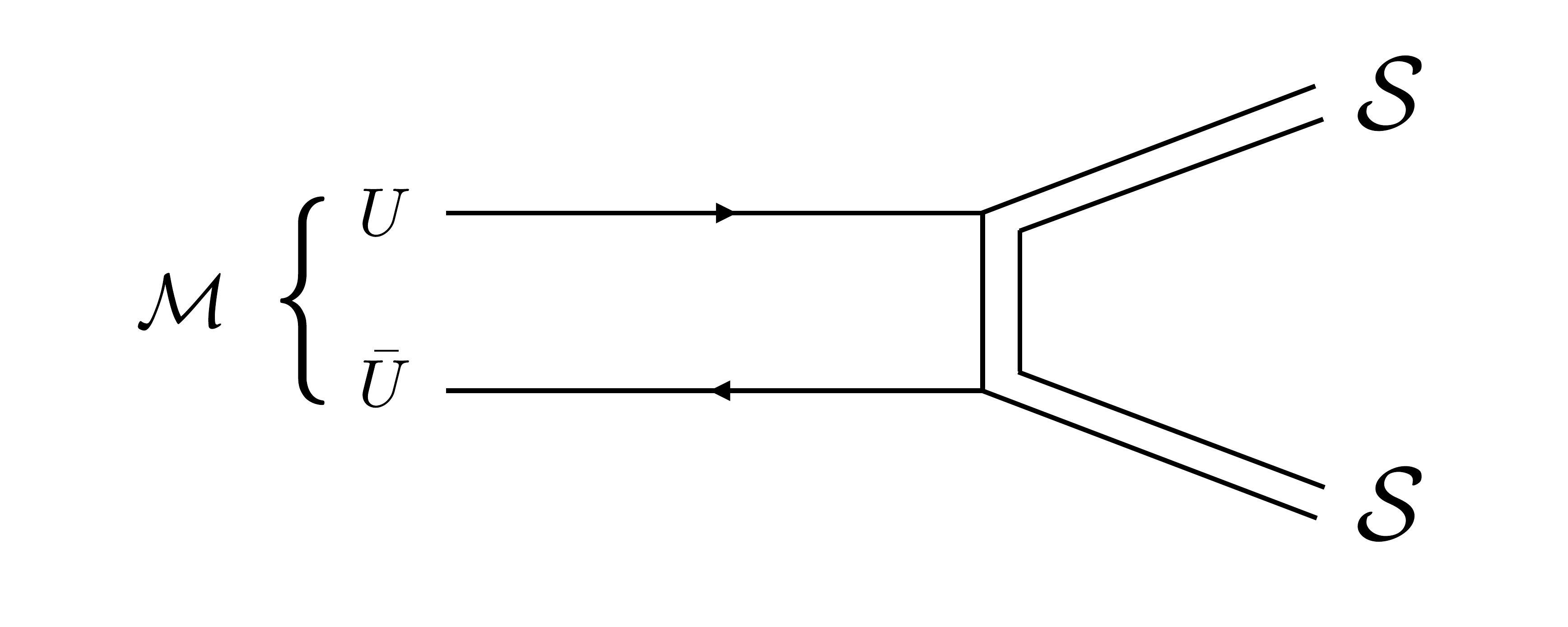}
 \end{minipage}
 \hspace{1cm}
 \begin{minipage}{.43\linewidth}
  \includegraphics[width=\linewidth]{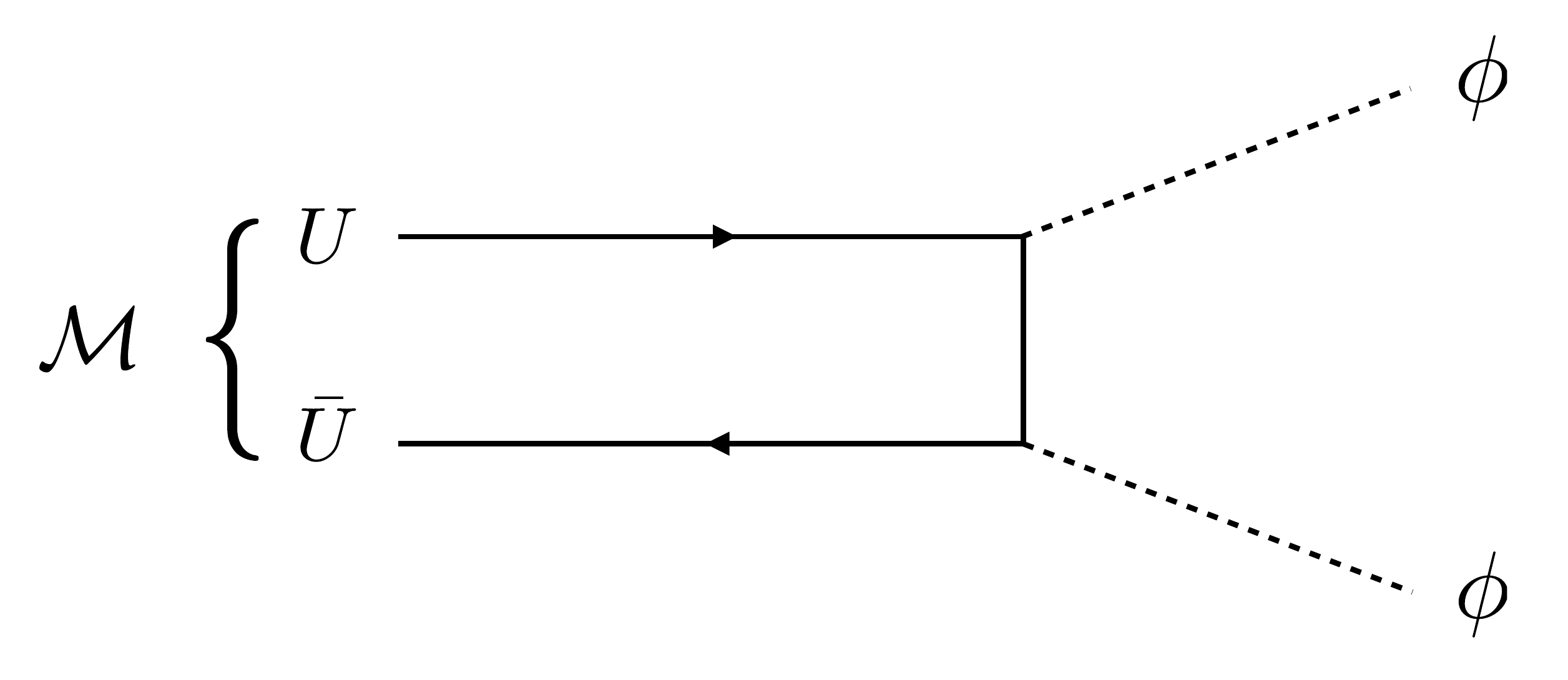}
 \end{minipage}
	\caption{The examples of the decay diagrams of ${\cal M}$ in which the quarks are annihilating.}
	\label{fig:decayM}
\end{figure}

As we have seen above (e.g. Fig.\,{\ref{fig:RT}}), the bound states shrink and get de-excited 
to the ground state once the temperature of the universe becomes much lower than $T\simeq T_c$.
Once the bound states stay in the ground state, they immediately decay into the glueballs
and the scalars $\phi$ (i.e. $a$'s and $\rho$'s) in which the heavy quarks annihilate microscopically 
(Fig.\,\ref{fig:decayM}).
The decay rate is given by the annihilation rate multiplied by the radial wave function 
of the ground state at around the origin,%
\footnote{The Bohr radius is of the order of $(\alpha_{N_c} M_U)^{-1}$.}
\begin{eqnarray}
\Gamma_{{\cal M}_0} \sim \frac{\pi\alpha_{{N_c}, g}^2}{M_U^2} \times (\alpha_{N_c} M_U)^3 \ .
\end{eqnarray}
Since this rate is much larger than the Hubble expansion rate, the mesons decay away very quickly. 
It should be also noted that 
the bound states spend a small fraction of their time as the ground state
even around $T\simeq T_c$. 
Thus, the mesons start to decay without waiting for complete de-excitation, as long as $\Gamma_{{\cal M}_0}\times \xi_1$
is larger than the Hubble expansion rate.
As a result, we find that the mesons decay away from the thermal bath immediately  for $T\lesssim T_c$.

\begin{figure}[tbp]
	\centering
  \includegraphics[width=.45\linewidth]{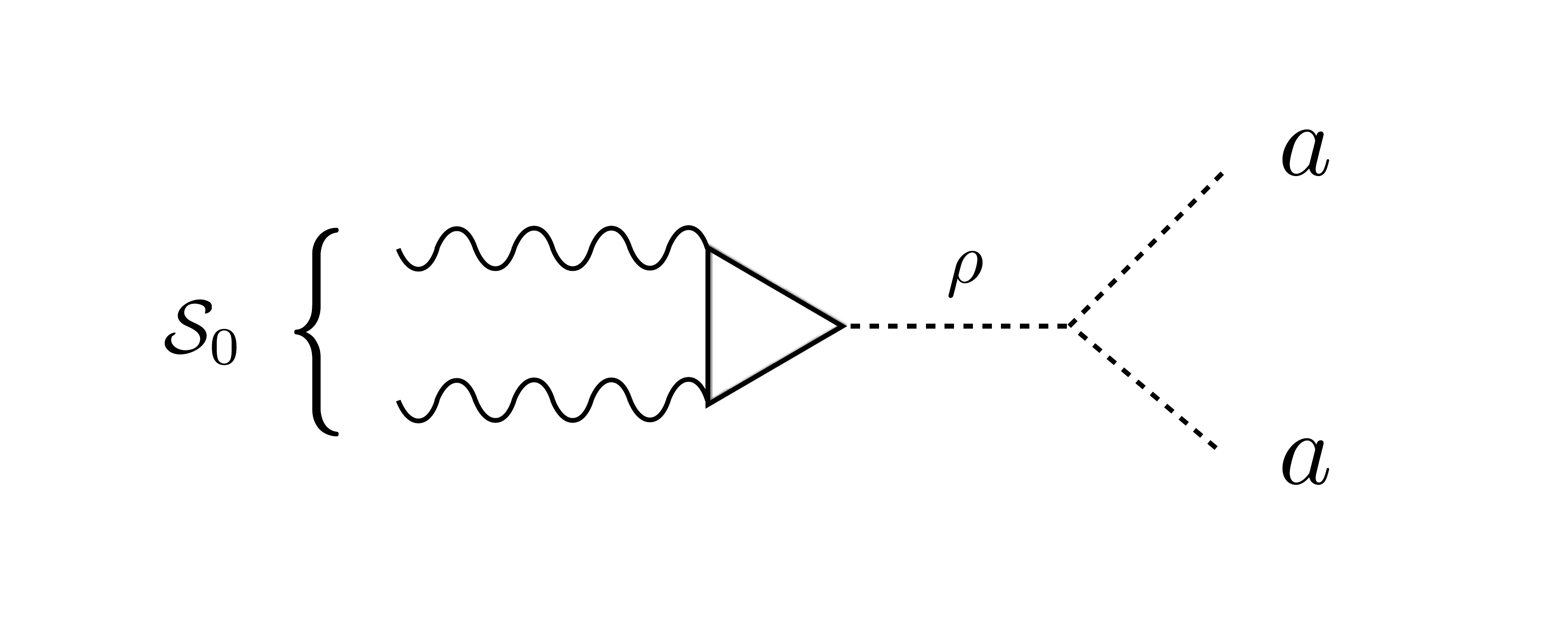}
	\caption{The examples of the decay diagrams of ${\cal S}$ through the mixing to $\rho$.
	In the triangle diagram, $U$ and $\bar{U}$ are circulating.}
	\label{fig:Sdecay}
\end{figure}

Excited glueball states  decay into lower-lying states immediately.%
\footnote{The masses of some low-lying states may be smaller that the twice of the mass of the ground state glueball.
Those states decay by emitting off-shell glueballs and have decay rates similar to the one of the ground state.}
The ground state $CP$-even glueball, ${\cal S}_0$, decays into a pair of the axions through the mixing to $\rho$ 
(see Fig.\,\ref{fig:Sdecay}).
The $CP$-odd glueball decays into a pair of ${\cal S}_0$ and an axion with a much higher rate.
The decay rate of the $CP$-even ground state glueball is roughly estimated by,
\begin{eqnarray}
\label{eq:GS0}
 \Gamma_{{\cal S}_0} &\sim& \frac{1}{8\pi}
\left(\frac{1}{4\pi}\right)^{2}
\left( \frac{\Lambda_{\rm dyn}}{f_a}\right)^2
\left( \frac{m_{\cal S}^2}{m_\rho^2}\right)^2 
 \frac{m_{\cal S}^3}{f_a^2}
\nonumber \\
& \sim & 10^{-12}\,{\rm GeV}
 \left(
 \frac{\Lambda_{\rm dyn}}{10^3\,{\rm GeV}}
 \right)^5
 \left(
 \frac{10^6\,{\rm GeV}}{f_a} 
 \right)^4
 \left( \frac{m_{\cal S}^2}{m_\rho^2}\right)^2 
 \left( \frac{m_{\cal S}}{\Lambda_{\rm dyn}}\right)^3 
 \ .
\end{eqnarray}
Here, the mixing angle between $\rho$ and ${\cal S}_0$ is estimated to be,
\begin{eqnarray}
{\varepsilon} \simeq \frac{1}{4\pi} \frac{\Lambda_{\rm dyn}}{f_a} \left( \frac{m_{\cal S}^2}{m_\rho^2}\right)\ ,
\end{eqnarray}
based on the Naive Dimensional Analysis\,\cite{Luty:1997fk,Cohen:1997rt}.
In terms of the cosmic temperature, the decay temperature of the glueball is roughly given by,
\begin{eqnarray}
\label{eq:Tdglue}
T_{{\cal S}_0} \simeq 10^{3}\,{\rm GeV}
 \left(
 \frac{\Lambda_{\rm dyn}}{10^3\,{\rm GeV}}
 \right)^{5/2}
\left( \frac{10^6\,{\rm GeV}}{f_a} \right)^2 
 \left( \frac{m_{\cal S}^2}{m_\rho^2}\right)
  \left( \frac{m_{\cal S}}{\Lambda_{\rm dyn}}\right)^{3/2} \ .
\end{eqnarray}
Thus, the glueballs also decay away immediately unless $\rho$ is very much heavier than $m_{\cal S}$. 

The massive glueballs decouple from the thermal bath when their annihilation into the axions 
freeze-out, which leaves the yield of the glueballs,
\begin{eqnarray}
Y_{\cal S} \sim \frac{x_F f_a^4}{M_{\rm PL} \Lambda_{\rm dyn}^3}\ ,
\end{eqnarray}
where we approximate $m_{\cal S} \simeq \Lambda_{\rm dyn}$.%
\footnote{Excited glueballs have much smaller yields.}
The relic glueballs would dominate the energy density of the universe at the temperature 
$T_{\rm dom} \simeq m_{\cal S} Y_{\cal S}$ if they are stable.
To avoid large entropy production by the decay of the glueballs, we  require 
so that $T_{{\cal S}_0} > T_{\rm dom}$.
We also require that ${\cal S}_0$ decays before the era of the Big-Bang Nucleosynthesis.%
\footnote{Even if $T_{{\cal S}_0}< T_{\rm dom}$, the present model provides a consistent
dark matter model as long as this condition is satisfied. 
In this case, the resultant dark matter density is further reduced than the one in the following 
estimation.}
Let us note here that ${\cal S}_0$ decays more efficiently
without requiring $m_{\rho} \ll {\cal O}(f_a)$ in the Higgs portal model discussed in the appendix \ref{sec:Higgs portal}.

\begin{figure}[tbp]
	\centering
  \includegraphics[width=.5\linewidth]{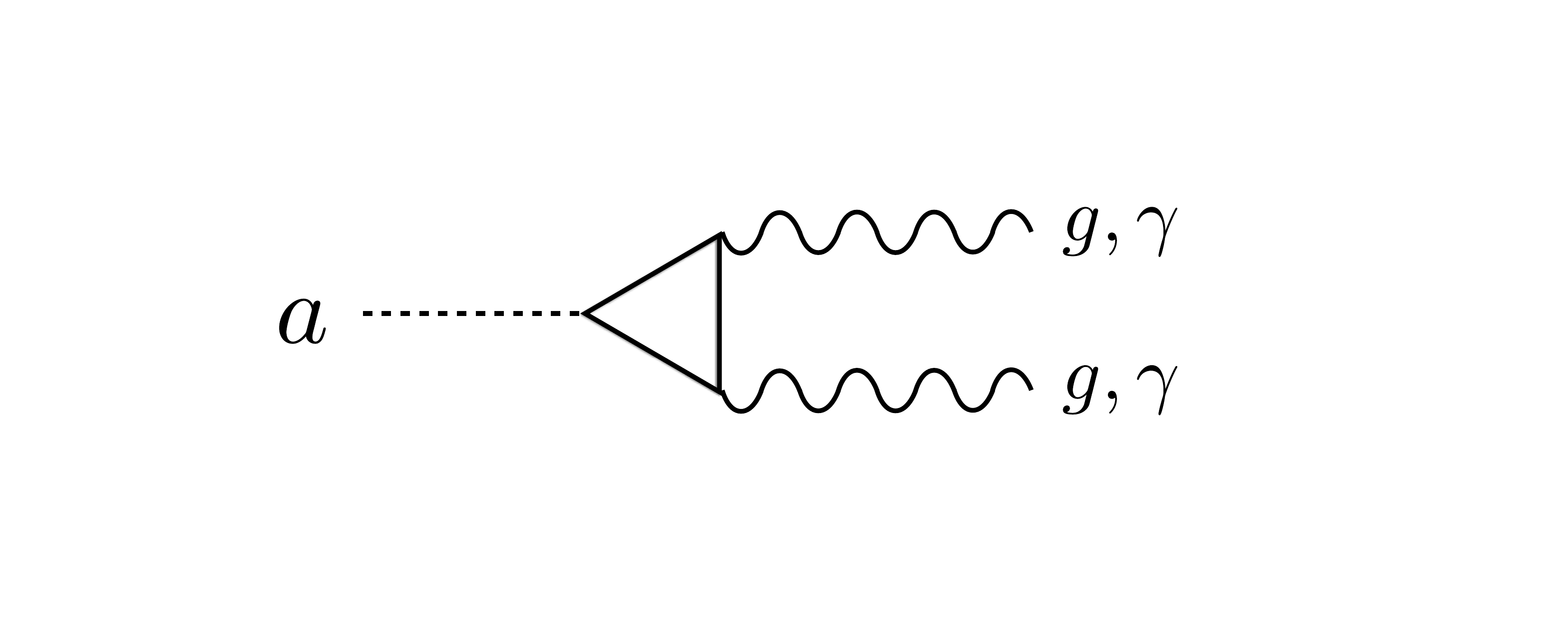}
	\caption{The examples of the decay diagrams of ${a}$ into the Standard Model gauge bosons.
	In the triangle diagram, $d'$ and $\bar{d}'$ are circulating.}
	\label{fig:adecay}
\end{figure}

Finally, the axion decays into the Standard Model particles via the anomalous coupling in Eq.\,(\ref{eq:anomaly})
(see Fig.\,\ref{fig:adecay}).
For $m_a \gtrsim {\cal O}(1)$\,GeV, the axion mainly decays into the QCD jets.
For $m_a \lesssim {\cal O}(1)$\,GeV, the axion decays into light hadrons through the mixing to the $\eta$ and $\eta'$ mesons 
in the Standard Model~\cite{Chiang:2016eav}.
It should be noted that the axion lighter than ${\cal O}(10-100)$\,MeV are excluded by  astrophysical constraints depending 
on the decay constant~\cite{Raffelt:2006cw}.
In our discussion, we assume $m_a \gtrsim {\cal O}(1)$\,GeV 
which is provided by the anomaly of $SU(N_c)$  (see Eq.\,(\ref{eq:axion})) or by other explicit breaking of the $U(1)_A$ symmetry if necessary.
Under this assumption, the axion also decays immediately at the temperature around $T\lesssim m_a$.

\subsection{Baryon Abundance}
Now, let us discuss the fate of the baryonic bound state.
Assuming a similar phenomenological potential for the quarks in the baryonic bound states,%
\footnote{Our assumption corresponds to the so-called the $\Delta$-law, where
the long-range potential is simply the sum of two-body potentials.
See Refs.~\cite{Richard:1992uk,Bali:2000gf} for more on phenomenological potentials for baryons.}
we expect that the baryons spend  most of their time as the excited states
and the typical size of the bound state is $R(T_c) \simeq {\cal O}(\Lambda_{\rm dyn}^{-1})$ at $T\simeq T_c$.
It should be noted that the baryons cannot decay away although they spend a small fraction of their time in the ground state
due to the $U(1)_B$ symmetry.

\begin{figure}[tbp]
	\centering
  \includegraphics[width=.45\linewidth]{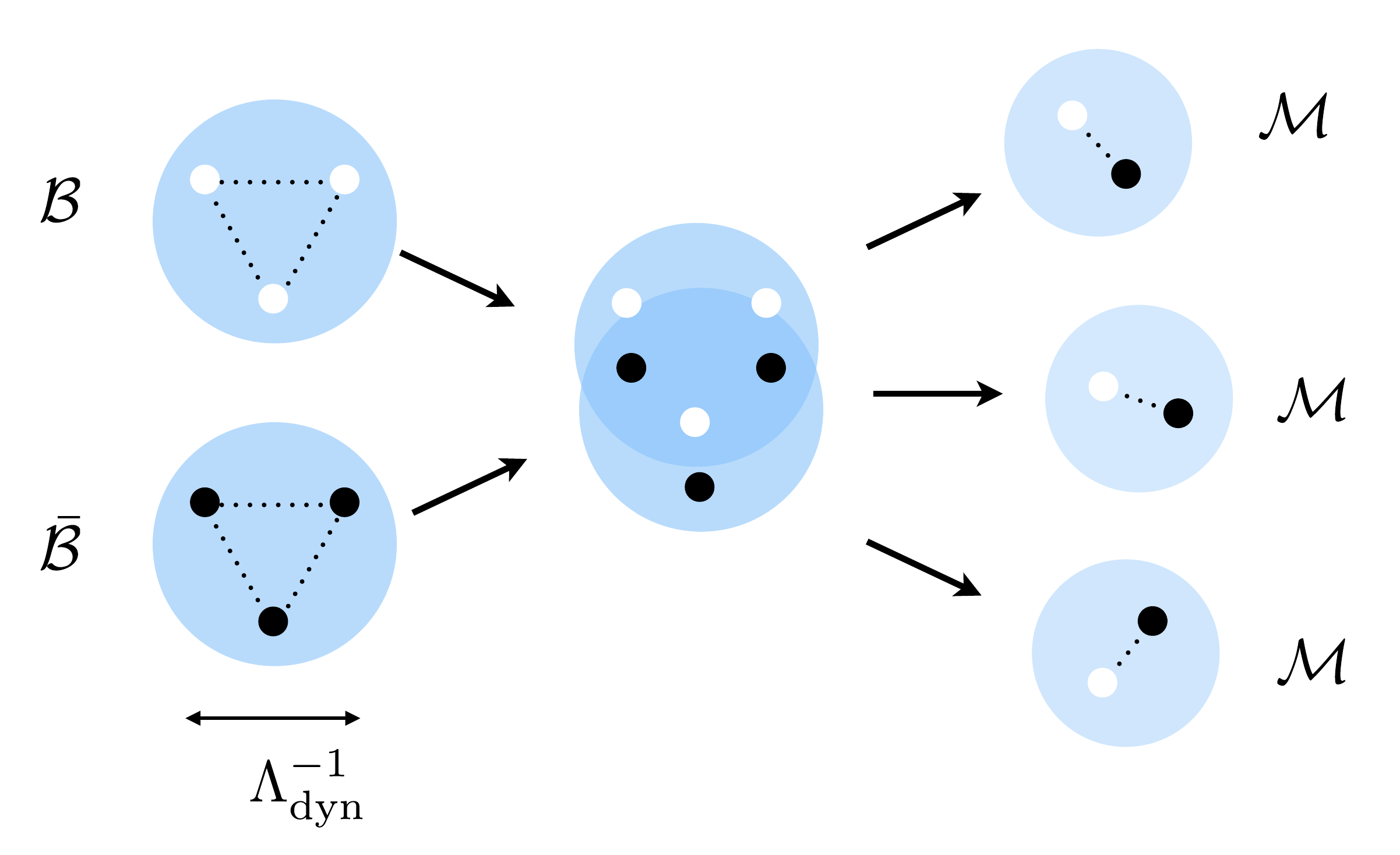}
	\caption{A schematic picture of the baryon annihilation into the mesons.
	The quarks stay in the overlapped region for a long time and they are reconnected to the mesons with ${\cal O}(1)$
	probability in each collision. 
	}
	\label{fig:BB}
\end{figure}

As a notable feature of the baryons, the baryons are able to annihilate into multiple mesons 
\begin{eqnarray}
\label{eq:Bannihilation}
{\cal B} + \bar{\cal B} \to {\cal M} + {\cal M} + {\cal M} + ({\cal S}) +  \cdots\  .
\end{eqnarray}
The cross section of this process is expected to be about a geometrical one,
\begin{eqnarray}
\sigma_B = A \pi R^2(T_c)\ ,
\label{eq:sigB}
\end{eqnarray}
where $A = {\cal O}(1)$.
In fact, as discussed in Ref.~\cite{Kang:2006yd}, 
the heavy quarks inside the bound states are moving very slowly, $v\sim \sqrt{\Lambda_{\rm dyn}/M_U}$
when the baryons are colliding.
Hence, the quarks stay in  overlap regions of the bound states for a long time,
 $\mit \Delta t \sim \sqrt{M_U/\Lambda_{\rm dyn}^3}$ in the collisions.
As a result, the quarks and anti-quarks are largely disturbed during the collision and they are well stirred.
Eventually, the quarks and the anti-quarks are reconnected so that the baryons are 
broken into the mesons with ${\cal O}(1)$ probability in each collision (see Fig.\,\ref{fig:BB}).
Once the annihilation into the mesons happens, the mesons in the final state immediately decay  into
glueballs as discussed in the previous section.

With the above annihilation cross section, the Boltzmann equation of the total number density of the baryon, $n_B$, is roughly given by,%
\footnote{Here, $\sigma_B$ denotes the annihilation cross section of each baryonic bound state,
which is roughly independent of the spins or any other internal degrees of freedom.
Thus, if there are $N_B$ species of the baryonic bound states, the Boltzmann equation of the number density of each species,
$n = n_B/N_B$, is given by,
\begin{eqnarray}
\dot{n} + 3 H n \simeq - N_B\times\vev{\sigma_Bv} n^2  \ .
\end{eqnarray}
}
\begin{eqnarray}
\label{eq:nB}
\dot{n}_B + 3 H n_B \simeq - \vev{\sigma_B v} n_B^2 \ .
\end{eqnarray}
By solving the Boltzmann equation, the number density of the baryons are reduced down to
\begin{eqnarray}
\frac{n_B}{s} \sim \left.\frac{H} {\langle{\sigma_B v}\rangle s}\right|_{T \simeq \Lambda_{\rm dyn}} 
\sim 3\times 10^{-16} \times A^{-1}
\left(\frac{M_U}{10^6\,{\rm GeV}}\right)^{1/2}
\left(\frac{\Lambda_{\rm dyn}}{10^3\,{\rm GeV}}\right)^{1/2}
\left(\frac{100}{g_*}\right)^{1/2}\ ,
\end{eqnarray}
leading to the relic abundance,
\begin{eqnarray}
\label{eq:Omega}
\Omega h^2  \sim 0.1\times \frac{N_c}{A} \left(\frac{M_U}{10^6\,{\rm GeV}}\right)^{3/2}
\left(\frac{\Lambda_{\rm dyn}}{10^3\,{\rm GeV}}\right)^{1/2}
\left(\frac{100}{g_*}\right)^{1/2}\ .
\end{eqnarray}
Here, the factor $N_c$ comes from the fact that the dark matter mass is $M_{B} \simeq N_c \times M_U$.
Therefore,
the observed dark matter density, $\Omega h^2 \simeq 0.1198\pm 0.0015$~\cite{Ade:2015lrj},
can be explained by the dark matter mass in the PeV range.

\begin{figure}[tbp]
	\centering
  \includegraphics[width=.4\linewidth]{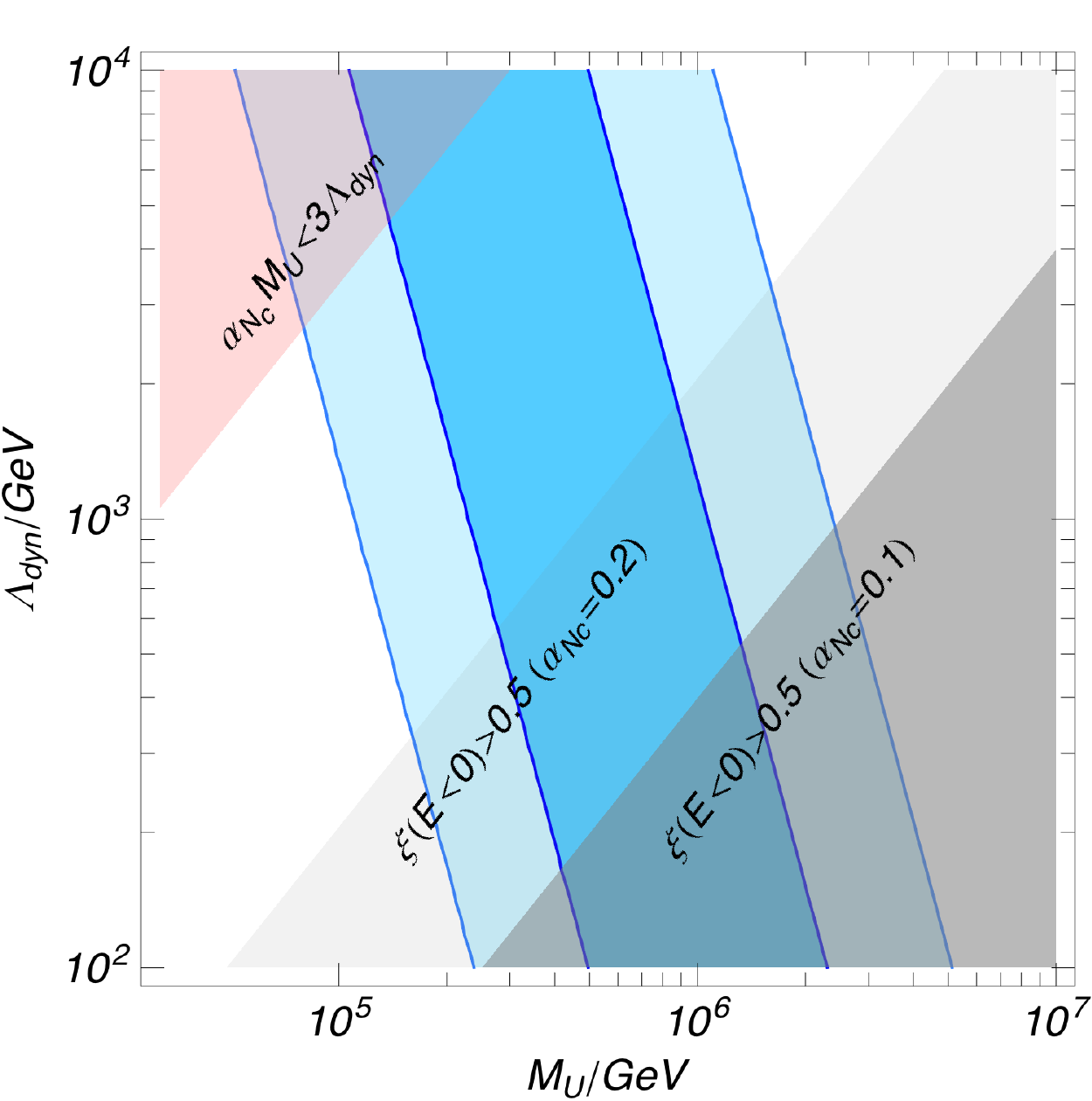}
	\caption{The parameter space which explains the observed dark matter density for $N_c = 3$.
	The dark matter mass is given by $M_{B} \simeq N_c \times M_U$.
	In the blue (light-blue) shaded region, the dark matter density in Eq.\,(\ref{eq:Omega}) reproduces the observed dark matter density for $A = 0.3$--$3$
        (for $A = 0.1$--$10$).
	In the gray shaded region, most of the bound states are in the negative energy region at around $T_c$ for $\alpha_{N_c} = 0.1$, and hence, the
	sizes of the bound states are rather small. (The light shaded region shows the same region for $\alpha_{N_c} = 0.2$.)
	In the pink shaded region, the gauge coupling constant is no more perturbative at the renormalization scale $\mu \sim \alpha_{N_c} M_U$. 
	}
	\label{fig:summary}
\end{figure}

In Fig.\,\ref{fig:summary}, we show the parameter space which can explain the observed dark matter 
density on the $(M_U, \Lambda_{\rm dyn})$ plane.
The blue shaded region explains the observed dark matter density for $N_c = 3$ with $A = 0.3$\,--\,$3$ in Eq.\,(\ref{eq:Omega}).
In the light-blue shaded region, the observed dark matter density is reproduced for $A = 0.1$\,--\,$10$.
In the gray shaded region, most of the bound states are in the negative energy region at around $T_c$ for $\alpha_{N_c} = 0.1$,
i.e. $\xi(E<0) = {\cal O}(1)$.
In such region, the sizes of the bound states are rather small at $T_c$, and hence, the annihilation cross section becomes smaller.
In the light-gray shaded region, we also show the same region for $\alpha_{N_c} = 0.2$.
The constraints from $\Gamma_{\rm ex} > H(T_c)$ lie below the gray shaded regions and hence are not shown.
In the pink shaded region, the gauge coupling constant becomes large at the renormalization scale $\mu \sim \alpha_{N_c} M_U$
where the one-loop running is no more reliable. 
It should be noted that the precise determination of the boundary of the allowed parameter space requires more detailed study
of the strong dynamics which goes beyond the scope of the present paper.
The figure shows that it is possible that the observed dark matter density is explained even for the dark matter mass 
$M_B \simeq N_c\times M_U$ with $M_U = {\cal O}(1)$\,PeV.

Let us emphasize here that the number density of the quarks   is conserved when the  baryons annihilate into the mesons.
The annihilation of the baryons just reconnects the quarks and anti-quarks inside the bound states.
The actual reduction of the number of quarks happens when the meson decays.
In this way, we can achieve a model of thermal relic dark matter with a mass lager than
the unitarity limit although no interaction violates the unitarity limit microscopically.

The consistency with the unitarity limit can also be understood in the following way~\cite{Griest:1989wd}.
When the dark matter particle has a radius of ${R = {\cal O}(\Lambda_{\rm dyn}^{-1}})$, the highest partial wave that contributes 
to the collision is 
\begin{eqnarray}
L_{\rm max} \sim M_U v\times R\ .
\end{eqnarray}
In this case, the annihilation cross section is bounded by the unitarity limit ,
\begin{eqnarray}
\sigma v \lesssim \sum_{L=0}^{L_{\rm max}}\frac{4\pi (2L+1)}{M_{U}^2 v} \sim \frac{4\pi L_{\rm max}^2}{M_{U}^2 v}  \sim 4\pi R^2 v \ .
\end{eqnarray}
This shows that  the geometrical cross section  in Eq.\,(\ref{eq:sigB}) is consistent with the unitarity limit.

\section{Conclusions and Discussions}
In this paper, we discussed a model with thermal relic dark matter where the dark matter mass
exceeds the so-called unitarity limit on the mass of  point-like particle dark matter.
In this model, the baryonic bound states are identified with dark matter, which 
possesses large radii when they are formed at the critical temperature around the the dynamical scale.
With the large radii, they annihilate into the mesons through a geometrical cross section.
The mesonic bound states decay into glueballs and axions which eventually decay into the 
Standard Model particles.
As a result, we found that thermal relic dark matter with a mass in the PeV range is possible, which
is beyond the usual unitarity limit.%
\footnote{It should be emphasized that the present paper does not require any entropy production to dilute the dark matter density.
For a heavy thermal relic dark matter scenario with entropy production see e.g. \cite{Berlin:2016vnh}.}

One caveat is that we assumed the same quark potential in the mesons and the baryons in our discussion.
If the binding energies of the baryons by the Coulomb potential are much larger than the mesons in Eq.\,(\ref{eq:En}), 
the size of the baryons at $T_c$ can be much smaller.
In this case, the baryon annihilation cross section is expected to be smaller than the one in Eq.\,(\ref{eq:sigB}),
and hence, the upper limit on the dark matter mass should be lower.
If the binding energies of the baryons are smaller than the mesons in Eq.\,(\ref{eq:En}), on the other hand, the
upper limit on the dark matter mass can be weaker.
To derive precise upper limit on the dark matter mass, we need to solve the strong gauge dynamics with heavy quarks precisely,
which is quite challenging with the current techniques.

In the model presented in this paper, we have the axion which couples to both the dark matter sector
and the Standard Model sector.
It is an interesting question whether the axion in the present model can play the role of the axion which solves 
the strong $CP$-problem by identifying $U(1)_A$ with the Peccei-Quinn symmetry\,\cite{Peccei:1977hh,Peccei:1977ur,Weinberg:1977ma,Wilczek:1977pj}.
Since the $U(1)_A$ symmetry is not only broken by the QCD but also by $SU(N_c)$ which possesses
its own $\theta$-term, it is apparently difficult for the axion in this model to solve the strong $CP$-problem. 
However, if the $SU(N_c  = 3)$ can be regarded as a counterpart of the QCD in a mirror copy 
of the Standard Model,%
\footnote{Here, we assume $Z_2$ exchange symmetry between the Standard Model and the copied sector,
which is broken spontaneously.}
the $\theta$ terms in $SU(N_c =3)$ and the QCD are aligned,
so that the axion in the present model might solve the strong 
$CP$-problem\,\cite{Rubakov:1997vp,Berezhiani:2000gh,Hook:2014cda,Fukuda:2015ana,Albaid:2015axa,Blinov:2016kte}.
Such a possibility will be discussed elsewhere.

Finally, let us comment on a possible phenomenological application of the present model.
In recent years, the IceCube experiment~\cite{Aartsen:2013jdh,Aartsen:2015zva,Aartsen:2015knd}
has reported the excess in the observed flux of extraterrestrial neutrinos
in the PeV range.
Dark matter with a mass in the PeV range is considered to be one of the attractive explanation of
the excess~\cite{Feldstein:2013kka,Barger:2013pla,Esmaili:2013gha}. 
For example, the excess can be accounted for by dark matter with spin $3/2$ and a mass $2.4$\,PeV
which decays into neutrinos via
\begin{eqnarray}
\label{eq:decay}
{\cal L} = \frac{1}{M_*} (\bar{L} iD_\mu H^c) \gamma^\nu \gamma^\mu\psi_\nu
\end{eqnarray}
for $M_* \simeq 5\times 10^{34}$\,GeV (corresponding lifetime of dark matter of ${\cal O}(10^{28})$\,s) \,\cite{Feldstein:2013kka}.
Here, $L$ and $H$ represent the lepton and Higgs doublets in the Standard Model
and $\psi_\nu$ is dark matter with spin $3/2$, respectively,

A serious drawback in the dark matter interpretation of the PeV neutrino flux is that its relic density cannot 
be explained by thermal relic density due to the unitarity limit.
As we have discussed, however, thermal relic density can be consistent with the observed 
dark matter even for PeV dark matter. 
In fact, $\psi_\nu$ can be identified with the baryons $N_c = 3$.%
\footnote{If the operator in Eq.\,(\ref{eq:decay}) is provided by a Planck suppressed operator 
of the quarks, $M_*$ is expected to be much larger than $M_* \simeq 5\times 10^{34}$\,GeV.
To provide appropriate $M_*$, we need further extension of the model
at the energy scale much larger than $M_U$ such as the emergence of  conformal dynamics. 
}
Therefore, the IceCube results can be interpreted by the decay of PeV {\it thermal relic dark matter}
in the present model.

\begin{acknowledgments}
This work is supported in part by Grants-in-Aid for Scientific Research from the Ministry of Education, Culture, Sports, Science, and Technology (MEXT), Japan, No. 25105011 and No. 15H05889 (M. I.); Grant-in-Aid No. 26287039 (M. I.) from the Japan Society for the Promotion of Science (JSPS); 
and by the World Premier International Research Center Initiative (WPI), MEXT, Japan (M. I.).
This work is also supported in part by the Department of Energy, Office of Science, Office of High Energy Physics, under contract No. DE-AC02-05CH11231 (K.~H.), by the National Science Foundation under grants PHY-1316783 and PHY-1521446 (K.~H.).
This work is also supported by IBS under the project code, IBS-R018-D1.
\end{acknowledgments}

\appendix
\section{Model with Higgs Portal}
\label{sec:Higgs portal}
In the main text, we assumed that the dark matter sector is connected to the Standard Model dominantly through the axion.
In this appendix, we consider an alternative model to connect the dark matter sector to the Standard Model sector 
via the Higgs portal.%
\footnote{In this model, the $U$-quarks annihilates not into $\phi$'s but into gluons and/or higgs at the perturbative freeze-out,
which does not affect the thermal history after the confinement.}

For that purpose, we introduce two additional flavors of the fermions in addition to the $U$-quarks, and assume that 
they form the doublet representation of the $SU(2)_L$ and have hypercharges of $\pm 1/2$.  of the Standard Model gauge groups.
We call the doublet quark $(U_H, \bar{U}_H)$ and  couple them to the Standard Model Higgs doublet $H$ via,
\begin{eqnarray}
{\cal L} = y H^\dagger U_H \bar{U}  +  y H \bar{U}_H U  + M_H \bar{U}_HU_H +  M_U \bar{U} U + h.c.
\end{eqnarray}
Here, $M_H$ is taken to be somewhat larger than $M_U$ so that they do not affect the properties
of the mesons and baryons discussed in the main text.
We do not  need to have a complex scalar field $\phi$ in this model.
$U_H$'s  decay into a pair of the Higgs doublet and a quark $U$.

By integrating out $U_H$, we obtain an effective coupling between the Higgs doublets and the 
$SU(N_c)$ gauge bosons,
\begin{eqnarray}
{\cal L} \sim \frac{\alpha_{N_c}y^2}{4\pi M_H^2}  H^\dagger H G_{N_c}\,G_{N_c}\ .
\end{eqnarray}
The Lorentz indices of the field strengths $G_{N_c}$ of $SU(N_c)$ should be understood.

The advantage of the model with the Higgs portal is the efficient decay of the lightest glueballs.
In fact, the above operator leads to an effective operator
\begin{eqnarray}
{\cal L} \sim\frac{1}{4\pi}  \frac{y^2 \Lambda_{\rm dyn}^3}{M_H^2}  H^\dagger H {\cal S}\ ,
\end{eqnarray}
which leads to a decay width,
\begin{eqnarray}
{\Gamma}_{{\cal S}_0} \simeq \frac{y^4}{8\pi}\left(\frac{1}{4\pi}\right)^2 \frac{\Lambda_{\rm dyn}^5}{M_H^4}\ .
\end{eqnarray}
Here, we again use the Naive Dimensional Analysis\,\cite{Luty:1997fk,Cohen:1997rt}.
Therefore, the decay width and the corresponding decay temperature of ${\cal S}_0$ 
can be as large as the ones in Eqs.\,(\ref{eq:GS0}) and \,(\ref{eq:Tdglue}) for $m_{\rho} \simeq \Lambda_{\rm dyn}$.
Thus, in the model with Higgs portal, the glueball decays efficiently without requiring $m_{\rho} \ll {\cal O}(f_a)$.

\section{Model with Hypercharge Portal}
\label{sec:hyper}
As another alternative model, we may consider an $SU(N_c = 3)$ model with two flavors $(U, \bar{U})$ and $(D,\bar{D})$
where $U$ and $D$ ($\bar U$ and $\bar D$) possess $U(1)_Y$ charges $2/3$ and $-1/3$ ($-2/3$ and $1/3$), respectively.
We assume that $U$ and $D$ have almost the same masses, 
\begin{eqnarray}
{\cal L} =   M \bar{U} U +  M \bar{D} D + h.c. \ ,
\end{eqnarray}
so that the model possesses an approximate global $SU(2)$ symmetry.

In this case, the light baryon states consist of an $SU(2)$ doublet baryons, 
\begin{eqnarray}
{\cal N} = (UDD,  UUD) \ ,
\end{eqnarray}
with a spin $1/2$ and an $SU(2)$ quadruplet baryons, 
\begin{eqnarray}
{\mit \Delta} = (DDD,  UDD ,  UUD,  UUU )\ ,
\end{eqnarray}
with a spin $3/2$.
Due to the spin-spin interaction, we expect that $\Delta$ is heavier than ${\cal N}$
by

\begin{eqnarray}
\label{eq:dM1}
{\mit \D}M_{{\cal N}- \mit \D} \sim \alpha_{N_c}^4 M  \ .
\end{eqnarray}
Furthermore, the neutral baryon ${UDD}$ is lighter due to the $U(1)_Y$ interaction, by,
\begin{eqnarray}
\label{eq:dM2}
{\mit \D}M \sim \alpha_{Y} \alpha_{N_c}M\ .
\end{eqnarray}
Therefore, in this case, the lightest baryon is expected to be $UDD$ in ${\cal N}$, which is neutral
under $U(1)_Y$ and can be identified with dark matter.%
\footnote{Due to the radiative corrections of $U(1)_Y$ gauge interactions,
the mass of the $D$ quark is expected to be smaller than that of the $U$ quark.
Here, we assume that the masses of $U$'s and $D$'s are finely tuned so that 
the mass differences between the baryons are mainly given by Eqs.\,(\ref{eq:dM1}) and (\ref{eq:dM2}).}

To  make unwanted charged particles in the dark matter sector decay, we introduce a light complex scalar field $s$ which
 has a hypercharge  $-1$ and the following coupling,
\begin{eqnarray}
{\cal L}  = y \,s\, U\bar{D} + h.c.
\end{eqnarray}
Though this interaction, the mesons decay into $s$'s (and glueballs) and the heavier baryons decay 
into the lightest baryon by emitting $s$'s.
Finally, $s$ decays into the Standard Model sector via, for example,
\begin{eqnarray}
{\cal L} = \frac{1}{M_*} \q^\mu s H \q_\mu H + h.c. 
\end{eqnarray}
where $M_*$ denotes a dimensionful parameter.%
\footnote{Since $s$ is charged under $U(1)_Y$, it should be heavy enough so that the 
constraints from the collider experiments are avoided.}

One might be interested in a model where $(U,D)$ and $(\bar{U},\bar{D})$ form the doublets of the 
$SU(2)_L$ gauge symmetry of the Standard Model with the hypercharges $1/6$ and $-1/6$, respectively.
In this case, the dark matter is again expected to be $UDD$ in ${\cal N}$, although the mass difference
between $UDD$ and $UUD$ is much smaller, ${\mit \D}M\simeq 347$\,MeV\,\cite{Cirelli:2005uq}.
Due to the couplings to the weak gauge bosons, the mesons and the heavier baryons immediately decay 
without introducing $s$.
It should be noted, however, that the direct detection experiments, the XENON\,100 \cite{Aprile:2012nq}
and the LUX  \cite{Akerib:2015rjg}, have put severe lower limit on the dark matter mass,
\begin{eqnarray}
 M_{\rm DM} > 3\mbox{--}5 \times 10^7\,{\rm GeV}\ .
\end{eqnarray}
Therefore, more suppression on the dark matter density is required for a consistent model (see Fig.\,\ref{fig:summary}).
For example, if $D(T_c) \gg M_U/F_{N_c}$ is achieved, we expect
further suppression of the dark matter density since the strings dominantly break-up 
and create a pair of the quarks and anti-quarks, and hence, most of the quarks are expected
to be trapped into mesons at $T_c$.
Such a possibility will be discussed elsewhere.

%

\end{document}